\documentclass[aps,prl,showpacs,twocolumn,floatfix,superscriptaddress]{revtex4}
\usepackage{graphicx} 
\usepackage{amsmath}
\usepackage{bm}
\usepackage{hyperref}
\usepackage{dcolumn}

\begin{document}

\title{Rydberg Spectroscopy in an Optical Lattice: Blackbody Thermometry for Atomic Clocks}

\author{Vitali D. Ovsiannikov}
\affiliation{Physics Department, Voronezh State University, Universitetskaya pl.\ 1, 394006, Voronezh, Russia}
\affiliation{Physics Department, University of Nevada, Reno,
Nevada, 89557, USA}

\author{Andrei Derevianko}
\affiliation{Physics Department, University of Nevada, Reno,
Nevada, 89557, USA}

\author{Kurt Gibble}
\affiliation{The Pennsylvania State University, University Park, PA, 16802 USA}

\begin{abstract}
We show that optical spectroscopy of Rydberg states can provide accurate {\em in situ} thermometry at room-temperature.  Transitions from a metastable state to Rydberg states with principal quantum numbers of  25 to 30 have 200 times larger fractional frequency sensitivities to blackbody radiation than the Strontium clock transition. We demonstrate that magic wavelength lattices exist for both Strontium and Ytterbium transitions between the metastable and Rydberg states. Frequency measurements of Rydberg transitions with $10^{-16}$ accuracy provide $10 \, \mathrm{mK}$ resolution and yield a blackbody uncertainty for the clock transition of $10^{-18}$.
\end{abstract}

\maketitle

\begin{figure}[h]
\begin{center}
\includegraphics*[scale=0.35]{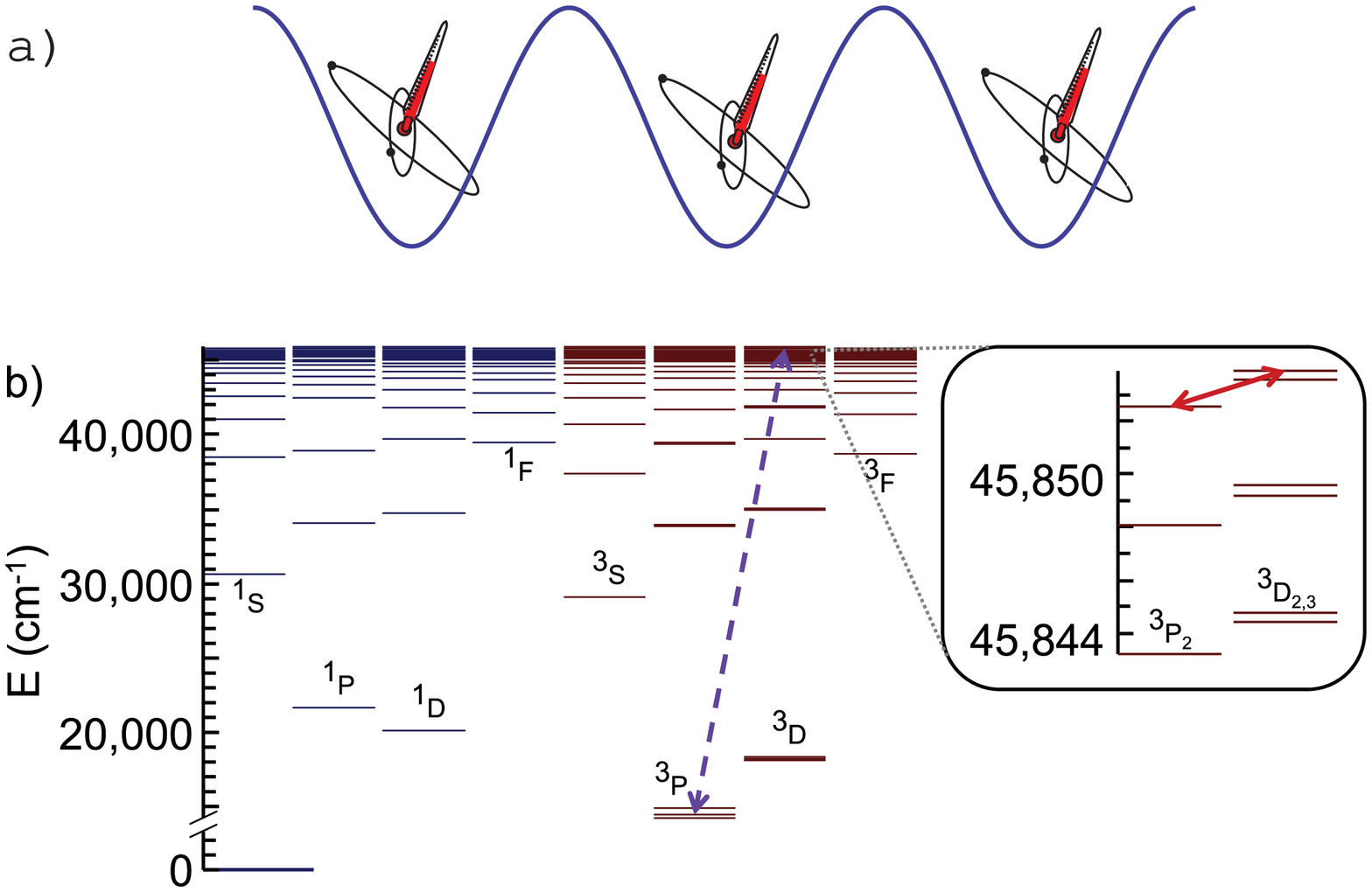}
\end{center}
\caption{(Color online) (a) Rydberg atoms trapped in an optical lattice can be sensitive thermometers. (b) Energy levels for Sr. Transitions from the metastable $^3P_0$ state to high Rydberg states (dashed) have a frequency sensitivity of 16 Hz/K, which can enable  a temperature accuracy of $\pm 10 \, \mathrm{mK}$. To achieve this accuracy using transitions between Rydberg levels (inset, solid), only a modest fractional frequency accuracy of $10^{-13}$ is required but, for Sr, the transition linewidths would require lines to be split by more than $10^{6}$. For $l \le 4$ states, transitions between $^3P_0$ and $^3D_1$ states have the largest sensitivity, $\approx 1Hz/K$; their precise energies are not known. }
\label{Fig:Setup}
 \end{figure}

Blackbody radiation (BBR) at room temperature limits the accuracy of optical and microwave atomic clocks. The root-mean-square electric field of BBR is $832\, \mathrm{V/m}$ at $300 \, \mathrm{K}$ and this produces problematic Stark shifts of atomic levels. Two currently promising candidates for future optical frequency lattice clocks are Sr and Yb\cite{KatTakPal03,LudZelCam08,LeTBaiFou06,MidLisFal10,PorDerFor04,LemLudBar09} (see review~\cite{DerKat11}). For Sr clocks, the fractional BBR frequency shift is $\delta \nu/\nu= 5.49 \times 10^{-15} (T/300\, \mathrm{K})^4$~\cite{PorDer06}, giving a sensitivity of $7.3 \times 10^{-17} \mathrm{K}^{-1}$. Thus, achieving an accuracy goal of $10^{-18}$ \cite{MidLisFal10} requires a temperature accuracy of 10 mK near 300 K. In current Cs clocks, the best temperature accuracies are $\pm 0.2 \mathrm{K}$, averaged over the relevant clock volume ~\cite{GerNemWey10}. Cs and Rb microwave clocks have a comparable BBR sensitivity to Sr, as does Yb; all are within factors of 3~\cite{MitSafCla10}. The BBR shift can be dramatically reduced by cooling clocks with liquid nitrogen, to near 77 K~\cite{Lev10,MidLisFal10}. This reduces the BBR shift by a factor of 200, but still, the required temperature uncertainty of $\pm 0.6$ K for $10^{-18}$ clock accuracy may be difficult~\cite{MidLisFal10}. For some applications, especially space clocks, cryogens may be prohibitive. One alternative being pursued is to use clock transitions that are fortuitously much less sensitive to BBR, such as the Al$^+$ ion, and the Cd, Hg, Mg, and Zn clock transitions~\cite{RosSchHum07,HacMiyPor08,MitSafCla10}. Here we show how to use transitions to Rydberg states for accurate {\em in situ} thermometry (Fig.~\ref{Fig:Setup}(a)). Rydberg states have large BBR Stark shifts, 200 times greater sensitivity to BBR than the Sr clock states. We also describe a magic wavelength lattice for Rydberg transitions for which the dipole approximation is not valid. Rydberg lattices may be important for a variety of applications, including quantum information and computation~\cite{SafWalMo10}. 


Optical lattice clocks have the potential to achieve unprecedented frequency stability. Optical lattices can naturally trap up to $10^6$ atoms, giving a high signal-to-noise ratio on an optical frequency transition of $10^{15}\, \mathrm{Hz}$ with a sub-Hertz width. A lattice at the magic wavelength~\cite{KatTakPal03} does not perturb the frequency of the clock transition and suppresses important systematic errors such as Doppler shifts. Here we show that applying this high-resolution spectroscopic capability to Rydberg atoms in a magic wavelength lattice can provide accurate thermometry. Highly excited Rydberg states have a BBR energy shift which asymptotes to that of a free electron, $\pi(k_B T)^{2}/3 c^3 \approx 2.4 \, \mathrm{kHz}$ at 300 K~\cite{CooGal80}. It corresponds to a temperature sensitivity of $16 \,\mathrm{Hz/K}$ and therefore a spectroscopic accuracy of 0.16 Hz can yield an {\em in situ} temperature uncertainty of $\pm 10 \, \mathrm{mK}$.

The BBR Stark shift is given by the dipole strength of the nearby transitions and their energies. All atoms have the same Rydberg spectrum for highly excited states ($n>30$ in Fig.~\ref{Fig:Setup}(b)). Since the transition energies are much less than the mean BBR photon energy, all high Rydberg states have the same energy shift so the frequencies of transitions between them have a negligible sensitivity to BBR. An exception is transitions between Rydberg states and excited inner shell states or multiple electron excitations ~\cite{GalSanSaf81}. However, these states are not Rydberg states and therefore have short lifetimes that do not provide a sufficiently precise temperature resolution.

Relatively low-lying
Rydberg states, where the energy of nearby transitions is comparable to the BBR photon energy, can give a large temperature sensitivity (Fig.~\ref{Fig:Setup}(b) inset, solid) because the energy splittings are slightly different for different angular momentum. We show that transitions from relatively low-energy metastable states to moderately high Rydberg states are better candidates (Fig. 1(b) dashed). Here the sensitivity arises because low-energy states, with their large energy splittings to all other states, have much smaller BBR shifts than Rydberg states. A metastable state requires less energetic photons to reach the Rydberg state than the ground state, making the laser technology easier and more reliable, and the metrology of the transition and BBR temperature more accurate. We next discuss transitions from the lowest metastable state to high Rydberg states and later transitions between intermediate Rydberg states.

\begin{figure}[h]
\begin{center}
\includegraphics*[scale=0.75]{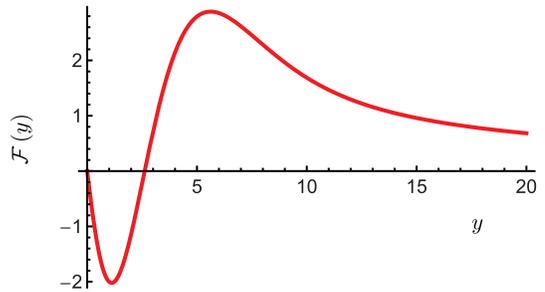}
\end{center}
\caption{(Color online) The Farley-Wing function~\protect\cite{FarWin81}is plotted versus normalized transition frequency $y= \omega_{n'n}/k_B T$. It gives the blackbody radiation shift of state $n$ due to $n'$, Eq.(\ref{dEFW}).}%
\label{Fig:Farley-Wing}%
\end{figure}

The general AC Stark shift of an atomic state due to BBR is
\begin{equation} \label{dE}
\Delta E_{n}^\mathrm{BBR}(T)=-\frac14\int\limits_0^\infty \mathcal{E}^2(\omega,
T)\alpha_{n}(\omega)d\omega
\end{equation}
in atomic units, where $\alpha_{n}(\omega)$ is the AC polarizability and the BBR spectral density
is
\begin{equation} \label{F}
\mathcal{E}^2(\omega, T)=\frac{8\omega^3}{\pi c^3\left(\exp(\omega/k_BT)-1\right)} \, .
\end{equation}
The BBR shift can be expressed as a sum over all dipole allowed transitions by integrating over the BBR spectrum
\begin{equation} \label{dEFW}
\Delta E_{n}^\mathrm{BBR}(T)=-\frac2{\pi c^3}\left(k_B T\right)^3
\sum\limits_{n'}  |\langle n'|z|nl\rangle|^2\mathcal F\left(
\frac{\omega_{n'n}}{k_B T} \right) \, .
\end{equation}
Here, the Farley-Wing function~\cite{FarWin81} $\mathcal{F}(y)$,
\begin{equation} \label{FFW}
\mathcal F\left(y\right)=-2y\int\limits_0^\infty \frac{x^3 dx}{(x^2-y^2)(e^x-1)},
\end{equation}
is plotted in Fig.~\ref{Fig:Farley-Wing}, where the integral is the Cauchy principal value.  For highly excited Rydberg states, nearby states give the largest dipole matrix elements, the energy splittings $\omega_{n'n}$ go to zero, and $\mathcal{F} \to - (\pi^2/3) y$. Summing over all states gives~\cite{FarWin81}:
\begin{equation} \label{dEFWas1}
\Delta E_{n}^\mathrm{BBR}(T)\approx \frac{\pi}{3 c^3}\left(k_B T\right)^2,
\end{equation}
which is 2.4 kHz at 300 K.

\begin{figure}[h]
\begin{center}
\includegraphics*[scale=0.5]{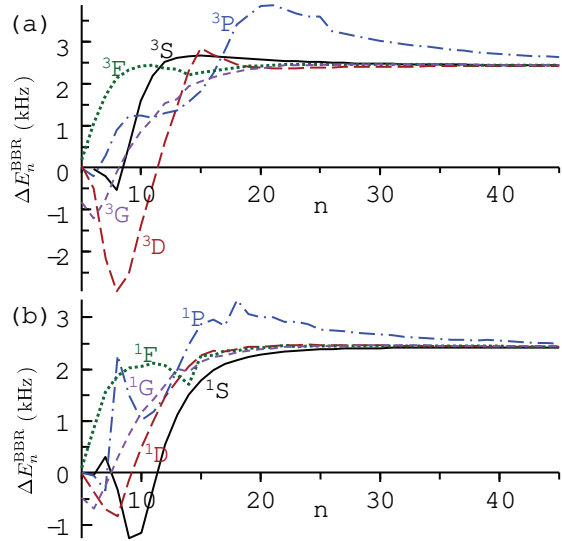}
\end{center}
\caption{(Color online) BBR Stark shifts of the (a) triplet and (b) singlet $5s n\ell $ Rydberg states of Sr. The shifts asymptote to 2.4 kHz for high principal quantum numbers $n$.}%
\label{Fig:BBRShifts}%
\end{figure}

We use a single-electron, model-potential method~\cite{ManOvsRap86} to calculate the BBR shifts for the lowest angular momentum states of Sr in Fig.~\ref{Fig:BBRShifts}. When the principal quantum number exceeds 25, all states have essentially the same BBR shifts. $D$ states with $n<9$ have negative BBR shifts because the transitions with the largest dipole moments have energies of order $k_B T$.

For the ground and the first excited states, the transition energies are much greater than $k_B T$. To lowest order,
\begin{equation} \label{dEFWgs}
\Delta E_{n}^\mathrm{BBR}(T)\approx -\frac{2 \pi ^3 \alpha_{n}(0) }{15 c^3} (k_B T)^4 \, ,
\end{equation}
which scales as $T^4$, instead of $T^2$ as for highly excited Rydberg states, since all transitions are far detuned. The difference of the static polarizabilities of the ground and excited clock states, $5s^2 \,^1\!S_0$ and $5s5p \,^3\!P_0$, give most of the sensitivity of the clock's frequency to BBR. The shifts are~\cite{PorDer06} -1.7 and -3.9 Hz respectively at $300 \mathrm{K}$, with sensitivities of $-0.011$ and $0.04\,\mathrm{K}^{-1}$. Thus, the BBR shifts of transitions from the ground or lowest metastable state to Rydberg states are dominated by the BBR shifts of the Rydberg states.

The advantage of probing $n>25$ states from a metastable state, as opposed to the ground state, is clear. The Sr transition wavelengths are shorter than 319 nm for the metastable state whereas they must be less than $219 \, \mathrm{nm}$ for the ground state. Higher $n$ are preferred because they have longer lifetimes, facilitating precise spectroscopic resolution, but this is tempered by a larger sensitivity to inter-atomic interactions and stray static electric fields. The Sr $5snd\,^3\!D_1$ states have significantly longer lifetimes than $L \neq 2$ states. The natural linewidth of the $5s25d\,^3\!D_1$ state is 1 kHz and its BBR broadening~\cite{CooGal80} is 2.5 kHz. A temperature accuracy of $10\,\mathrm{mK}$ requires the transition frequency to be known to $5 \times 10^{-5}$ of the linewidth, well less than the routine $10^{-6}$ line-splitting of microwave atomic clocks ~\cite{WynWey05}. The fractional frequency accuracy must be $1.7 \times 10^{-16}$ to allow $10^{-18}$ accuracy of the clock transition's BBR shift, giving an accuracy leverage of more than 100.

The DC Stark shifts of Rydberg levels from stray electric fields is a systematic error for BBR thermometry that has to be evaluated experimentally.  Asymptotically, the static polarizability of Rydberg states scales as $n^7$. The $5snd\,^3\!D_1$ state has a scalar polarizability of $-100 \, \mathrm{Hz \,  m^2/V^2}$ for $n=25$ and $-440 \, \mathrm{Hz \, m^2/V^2}$ for $n=30$.  Thus, by using more than one transition in this range, both the temperature and the magnitude of any stray electric fields can be determined.

Inter-atomic interactions between highly excited Rydberg states are large, even useful for Rydberg blockades~\cite{SafWalMo10}. For large energy shifts, as in effective blockades, the interactions are $R^{-3}$ where $R$ is the distance between atoms. Here, the high requisite precision dictates that the atoms interact weakly, and therefore the interactions are small compared to the Rydberg spacings (van der Waals regime) and the interactions scale as $R^{-6}$ and $n^{11}$. This limits the maximum viable $n$. For $n=25$, the mean spacing must be $4 \, \mathrm{\mu m}$ for a 1 Hz shift, which could be satisfied in a three dimensional lattice with unity occupation and a large lattice spacing. If the lattice volume is $(100 \, \mathrm{\mu m})^3$, more than $10^4$ atoms can be trapped. The shot-noise limited signal-to-noise would be 100, enabling a frequency resolution of 0.16 Hz in less than $10^3$ measurements of a $3.5 \, \mathrm{kHz}$ wide transition.


A magic-wavelength optical lattice is needed for Rydberg transitions just as it is for the high-accuracy spectroscopy of the clock transition. A lattice slashes the systematic errors from photon recoils and Doppler shifts. However, the size of Rydberg atoms can easily be larger than the periodicity of optical lattices~\cite{YouKnuAnd10}. Therefore, the dipole approximation may not be valid and Rydberg atoms could be untrapped. Here we show that Sr and Yb magic wavelength lattices do exist for Rydberg transitions, with $n$ as large as 40, from the $^3\!P_0$ metastable states.


We begin with the full interaction potential for the electromagnetic field,
\begin{equation} \label{Vint}
V(\mathbf r_e, t)= \frac{\mathbf A^2(\mathbf r_e, t)}{2c^2}-\frac1c (\mathbf A(\mathbf r_e, t)\cdot
\hat{\mathbf p}_e),
\end{equation}
where $\hat{\mathbf p}_e$ and $\mathbf{r}_e$ are the momentum operator and the coordinate of the Rydberg electron. The vector potential of a standing wave is, in the Coulomb gauge
\begin{equation} \label{A}
\mathbf A(\mathbf r, t)=-\frac{2c \mathcal{E}_0}{\omega}\mathbf
e_z\sin\left[k_m (X_0+x_e)\right]\sin(\omega_m t),
\end{equation}
where we separate out the nuclear coordinate $X_0$.
Breaking the atom-lattice interaction (\ref{Vint}) into two terms is particularly useful for Rydberg states. The dominant contribution is given by the first term if there is not an accidental resonance for the lower state. The second term only gives a correction, smaller by approximately $(n^2\omega_m)^{-2}\ll 1$. Thus, the Stark shift of a Rydberg transition is
\begin{eqnarray} \label{Starknl}
\delta E_{n} &=&
\frac{\mathcal{E}_0^2}{\omega_m^2}\left[\sin^2(k_m X_0)(1-2\langle
n|\sin^2(k_m x_e)|n\rangle)\right.
\nonumber\\
 &+&\left.\langle n|\sin^2(k_m x_e)|n\rangle\right].
\end{eqnarray}
For the metastable $^3\!P_0$ state, the dipole approximation is well satisfied giving the familiar Stark shift
\begin{equation} \label{Stark3P0}
\delta E_{^3\!P_0}
=-\mathcal{E}_0^2\alpha_{^3\!P_0}(\omega_m)\sin^2(k_mX_0).
\end{equation}
Comparing (\ref{Starknl}) and (\ref{Stark3P0}), we see that the lattice potential for the nuclear position terms $(X_0)$ is the same if
\begin{equation} \label{magic}
\alpha_{^3\!P_0}(\omega_m)=-\frac1{\omega_m^2}\left[1-2\langle
n|\sin^2(k_mx_e)|n\rangle\right].
\end{equation}
Thus, the polarizability of the metastable state has to be negative, $\alpha_{^3\!P_0}(\omega_m)<0$, so the magic wavelength must be blue-detuned from a metastable state's dipole resonance. The atoms are then confined at the intensity minima of a repulsive lattice.

The matrix element in the right-hand sides of (\ref{Starknl}) and  (\ref{magic}) can be evaluated analytically in the limit of small ($k_ma_n\ll 1$) and large ($k_ma_n\gg 1$) Rydberg orbits, of radius $a_n\propto n^2$:
\begin{itemize}
\item[(i)]
{$\langle n|\sin^2(k_mx_e)|n\rangle\approx
\frac13 k_m^2 \langle n|r^2|n\rangle\approx\frac56\ k_m^2 n^4 $ ($k_ma_n\ll 1)$
This correction is much less than one when $\lambda>1\,\mu m$ and $n <40$.}
\item[(ii)]{ $\langle n|\sin^2(k_mx_e)|n\rangle\approx 1/2$ ($k_ma_n\gg 1$),
In this limit, the first term in the square brackets in (\ref{Starknl}) goes to zero. The second term is independent of the nuclear position so, when the Rydberg orbit is much larger than the lattice wavelength, $k_ma_n\gg 1$, the Rydberg atom cannot be trapped in a lattice. However, if the lattices beams do not counterpropagate, the lattice spacing can be arbitrarily large - $k_m$ may be much less than $\omega_m$.}
\end{itemize}

Eq.~(\ref{magic}) gives a smooth dependence of the magic frequency on $n$. While the position independent term can be significant, it is supressed for large lattice spacings. For $n=25$ and a $4 \, \mathrm{\mu m}$ spacing, this term is $5.6 \times 10^{-4} \mathcal{E}_0^2 / {\omega_m^2}$. We note that higher-order multipoles (magnetic dipole M1 and electric quadrupole E2) \cite{TaiYudOvs08,KatHasIl'09} give a similar atom-position-independent term in (\ref{Stark3P0}) for the metastable state. These are negligible in comparison with those for the Rydberg state, which are automatically included in the form of (\ref{Vint}).

For Sr atoms, a range of magic wavelengths exists from 2379 nm (for $n = 40$) to 2392 nm (for $n = 15$). Here the polarizabilities range from $120$ to $129\, \mathrm{kHz/(kW/cm^2)}$. Calculations for Yb give magic wavelengths of 1142 nm ($n = 40$) to 1209 nm ($n = 15$), and polarizabilities from $18.8$ to $32.8\, \mathrm{kHz/(kW/cm^2)}$, for two-photon transitions to $6snp\,^3\!P_0$ states (see table \ref{t1}). A 3D lattice generally produces problematic vector and tensor light shifts. One way to control these is to use three pairs of linearly polarized "independent" beams that have slightly different frequencies, which gives an effective linear polarization throughout the lattice ~\cite{Winoto99}.

\begin{table}
\caption{\label{t1} Magic optical lattice wavelengths $\lambda_m$ for Yb transitions between the metastable $6s6p \,^3\!P_0$ and Rydberg states $6snp\,^3\!P_0$, $15\leq n \leq 40$. The corresponding polarizability $\alpha(\omega_m)$ gives the lattice depth, $U_{lat}=\alpha(\omega_m)I_\mathrm{lat}$ and we also list two-photon transition wavelength $\lambda_i$.}
\begin{center}
\begin{tabular}{ccdd}\hline
$n$&$\lambda_m$ &
\multicolumn{1}{c}{$\alpha(\omega_m)$}
&\lambda_i\\
& nm &
\multicolumn{1}{c}{
$\mathrm{kHz/(kW/cm^2)}$}& \mathrm{nm}\\
\hline
15&1209&$32.8$&$620.2$ \\
$20$&$1207$&$32.2$&$611.1$\\
$25$&$1203$&$31.1$&$607.8$\\
$30$&$1194$&$28.8$&$606.2$\\
$35$&$1178$&$25.1$&$605.3$\\
$40$&$1142$&$18.8$&$604.8$\\
\hline
\end{tabular}
\end{center}
\end{table}

BBR thermometry can also be performed with transitions between Rydberg levels.  The energy shifts in Fig.~\ref{Fig:BBRShifts}(a) show that there are large differences around $n=10$, where the energy splittings are comparable to $k_B T$ at 300K. However, the transition linewidths in this region are broad, of order 100 kHz, limiting the resolution. Nonetheless, a difference in sensitivity, albeit smaller, extends up to high $n$ and a number of transitions are sensitive. For example, at $n=40$, the BBR shifts and sensitivities for the $^3\!P_0$ and $^3\!D_1$ are (2,713 Hz, 17.06 Hz/K) and
(2,415 Hz, 16.14 Hz/K). The natural linewidths are 8,334 and 233 Hz, and the BBR broadenings are 1.9 kHz.  To achieve 10 mK temperature resolution, the line has to be split by a challenging factor of more than $10^6$. The advantage is that the transition frequencies are small, less than 60 GHz, and therefore require an effortless fractional frequency accuracy of only $1 \times 10^{-13}$. Because the linewidth is large, and because neither state is metastable, optical transitions from the metastable $^3\!P_0$ to Rydberg states appear more promising for Sr.

To summarize, blackbody radiation at room temperature poses a limit to the accuracy of Sr and Yb optical-frequency atomic clocks. Transitions from a metastable state to low-lying Rydberg states, $n=25-30$, have fractional frequency sensitivities to blackbody radiation that are 200 times larger than the Sr clock transition. {\em In situ} measurements of these Rydberg transition with an accuracy of $10^{-16}$ would give the temperature to $\pm 10 \, \mathrm{mK}$ and enable clock accuracies of $10^{-18}$. We show that magic wavelength lattices exist near 2.3 and 1.2 $\mu\mathrm{m}$ for these Rydberg transitions in Sr and Yb. Systematic errors such as Stark shifts from patch electric fields can be evaluated by probing several Rydberg transitions. Interactions between atoms limit the maximum density to less than $10^{10}\, \mathrm{cm}^{-3}$ for the Rydberg spectroscopy, which is still high enough to give sufficient signal-to-noise. While transitions between two Rydberg levels around $n=40$ could also be used for thermometry, the temperature sensitivity for these would require a highly accurate splitting of the transition linewidths, better than $10^{6}$. Rydberg thermometry may be particularly useful for clock applications where cryogens are prohibitive, including optical-frequency space clocks.

We thank M. Ahmed and H. Gharibnejad for assistance and acknowledge financial support from RFBR (VDO, No. 11-02-00152-a), the NSF (AD and KG),  the hospitality and support of the University Nevada, Reno (VDO and AD) and Penn State (KG).

We note that another theoretical analysis of trapping Rydberg states in lattices, to generate atomic entanglement, has recently appeared ~\cite{MukMilNat11}.


\begin{thebibliography}{10}

\bibitem{KatTakPal03}
H.~Katori, \emph{et~al.}, Phys.\ Rev.\ Lett. \textbf{91}, 173005 (2003).

\bibitem{LudZelCam08}
A.~D. Ludlow, \emph{et~al.}, Science \textbf{319}, 1805 (2008).

\bibitem{LeTBaiFou06}
R.~{Le Targat}, \emph{et~al.}, Phys. Rev. Lett. \textbf{97}, 130801 (2006).

\bibitem{MidLisFal10}
T.~{Middelmann}, \emph{et~al.} IEEE Trans. Instr. Meas. \textbf{60}, 2550 (2011).

\bibitem{PorDerFor04}
S.~G. Porsev, A.~Derevianko, and E.~N. Fortson, Phys. Rev. A \textbf{69},
  021403(R) (2004).

\bibitem{LemLudBar09}
N.~D. Lemke, \emph{et~al.}, Phys. Rev. Lett. \textbf{103}, 063001 (2009).

\bibitem{DerKat11}
A.~Derevianko and H.~Katori, Rev. Mod. Phys. \textbf{83}, 331 (2011).

\bibitem{PorDer06}
S.~G. Porsev and A.~Derevianko, Phys. Rev. A \textbf{74}, 020502 (2006).

\bibitem{GerNemWey10}
V.~Gerginov, \emph{et~al.}, Metrologia \textbf{47}, 65  (2010).

\bibitem{MitSafCla10}
J.~Mitroy, M.~S. Safronova, and C.~W. Clark, J. Phys. B \textbf{43}, 202001
  (2010).

\bibitem{Lev10}
F.~Levi \emph{et~al.}, IEEE Trans. on Ultrason. Ferro. Freq. Contr.
  \textbf{57}, 600 (2010).

\bibitem{RosSchHum07}
T.~Rosenband, \emph{et~al.}, Phys. Rev. Lett. \textbf{98}, 220801 (2007).

\bibitem{HacMiyPor08}
H.~Hachisu, \emph{et~al.}, Phys. Rev. Lett. \textbf{100}, 053001 (2008).

\bibitem{SafWalMo10}
M.~Saffman, T.~G. Walker, and K.~M\o{}lmer, Rev. Mod. Phys. \textbf{82}, 2313
  (2010).

\bibitem{CooGal80}
W.~E. Cooke and T.~F. Gallagher, Phys. Rev. A \textbf{21}, 588 (1980).

\bibitem{GalSanSaf81}
T.~F. Gallagher, \emph{et~al.}, Phys. Rev. A \textbf{23}, 2065 (1981).

\bibitem{FarWin81}
J.~W. Farley and W.~H. Wing, Phys. Rev. A \textbf{23}, 2397 (1981).

\bibitem{ManOvsRap86}
N.~L. Manakov, V.~D. Ovsiannikov, and L.~P. Rapoport, Phys. Rep. \textbf{141},
  319 (1986).

\bibitem{WynWey05}
R.~Wynands and S.~Weyers, Metrologia \textbf{42}, S64 (2005).

\bibitem{LodWesLem09}
J.~Lodewyck, P.~G. Westergaard, and P.~Lemonde, Phys. Rev. A \textbf{79},
  061401 (2009).

\bibitem{YouKnuAnd10}
K.~C. Younge, \emph{et~al.}, Phys. Rev. Lett. \textbf{104}, 173001 (2010).

\bibitem{TaiYudOvs08}
A.~V. Taichenachev, \emph{et~al.}, Phys. Rev. Lett. \textbf{101}, 193601
  (2008).

\bibitem{KatHasIl'09}
H.~Katori, \emph{et~al.}, Phys. Rev. Lett. \textbf{103}, 153004 (2009).

\bibitem{Winoto99}
S. L.Winoto \emph{et~al.}, Phys. Rev. A \textbf{59}, R19 (1999).


\bibitem{MukMilNat11}
R.~{Mukherjee}, \emph{et~al.}  (2011), arXiv:1102.3792.


\end{thebibliography}

\end{document}